\documentclass[twocolumn,english,aps, pre]{revtex4}
\usepackage[T1]{fontenc}
\usepackage[latin9]{inputenc}
\usepackage{textcomp}
\usepackage{amstext}
\usepackage{graphicx}
\usepackage{epstopdf}



\usepackage{babel}
\begin{document}

\title{Loss Factor of Supercooled Water at the Frequencies of 11$\ldots$180~GHz}

\author{G.S. Bordonskiy, A.O. Orlov}

\affiliation{Institute of Natural Resources, Ecology and Cryology SB RAS, 16 Nedorezova,
p/b 1032, 672002 Chita, Russia}

\email{lgc255@mail.ru}

\begin{abstract}
The loss factor of supercooled water at the frequencies $11\ldots180$~GHz has been measured. 
A measuring technique has been proposed, in which wetted nanoporous silicate materials, 
silica gels, with the mean diameter of the pores being $6-9$~nm, were used to obtain deeply 
supercooled water. 
Results have been obtained for the loss factor of supercooled water, close to volume water 
for its properties, when cooled down to $-45\,^{\circ}$C. 
To ascertain the mechanism of pore water losses, measurements have been made in the range of 
temperatures $0\ldots-90\,^{\circ}$C. 
The results obtained have demonstrated the existence of significant excessive losses at the 
temperatures below $-30\,^{\circ}$C, compared to the results of computations based on the 
known models. 
To allow mathematical description of the increment loss factor, a new addend has been 
introduced as a sum of two Gaussian functions in the formula described in [T. Meissner, 
F. J. Wentz, IEEE Trans. Geosci. Remote Sens. 2004. vol. 42, p. 1836]. 
One of these functions has the extremum near $-45\,^{\circ}$C, and the second one has the 
extremum in the range of $-60\ldots-70\,^{\circ}$C. 
Additional attenuation at $-45\,^{\circ}$C is supposed to be connected with the second 
critical point of water. 
Attenuation with the center in the range of temperatures $-60\ldots-70\,^{\circ}$C is 
determined by the emergence of conductive films at the boundary between the hard matrix 
and ferroelectric ice~0. 
This modification is a transitional form to ice~Ih or ice~Ic and is formed at the temperature 
below $-23\,^{\circ}$C. 
\end{abstract}

\maketitle

\section*{Introduction}

The knowledge of the dielectric parameters of supercooled metastable water is required for 
solving the problems of transfer of radiation in atmospheric aerosol, as well as in the 
frozen Earth cover. 
In a number of studies, models have been proposed for finding complex dielectric permittivity 
of water microwaves in the range of temperatures $0\ldots-45\,^{\circ}$C \cite{Ell2007, MWen2004, Ros2015, KDY2016, CT2011}. 
These models were based on a limited amount of experimental data \cite{BCS1982, HA1978, RThAWMK1997}, especially relating 
to the temperature range $-20\ldots-45\,^{\circ}$C. 
For example, in \cite{BCS1982} the lowest temperature at which measurements were made was 
$-18\,^{\circ}$C, and it was obtained in the micro emulsion of water. 
Different formulae presented in \cite{Ell2007, MWen2004, Ros2015, KDY2016, CT2011} demonstrate significant differences in the loss 
factor values ($\varepsilon^{\prime\prime}$) at the temperatures below $-20\,^{\circ}$C.
Therefore, experiments were made relating to finding this value in a broad range of 
frequencies and at the temperatures $-20\dots-45\,^{\circ}$C.

The main challenge in finding $\varepsilon^{\prime\prime}$ is to obtain the sufficient amount 
of supercooled water required for making the experiments. 
At the same time, it is known that deep supercooling of water is possible in porous bodies 
with nanopores \cite{SchKF2001, LCh2012}. 
In \cite{MatRC2010}, it was proposed to use such media for microwave measurements of the parameters of 
supercooled water. 
This idea was implemented in \cite{BOSch2017, BOKh2017}, in which silicate materials were used as porous media. 
The measurements of $\varepsilon^{\prime\prime}$ were made at certain frequencies and within 
the relatively narrow frequency ranges up to temperatures $\sim-90\,^{\circ}$C. 
The main result of these studies consists in the fact that, as opposed to \cite{MWen2004}, in the range 
of temperatures from $-30$ to $-70\,^{\circ}$C, redundant absorption of microwave radiation 
is observed. 
However, in \cite{MWen2004} the calculations of $\varepsilon^{\prime\prime}$ below $-45\,^{\circ}$C were 
not planned, as $\varepsilon^{\prime\prime}=0$ (at $-45\,^{\circ}$C). 
Therefore, in comparing the calculations with the results of this study, the loss factor was 
taken to be zero for temperatures below $-45\,^{\circ}$C. 
The studies performed \cite{BOSch2017, BOKh2017} resulted in obtaining approximations of the coefficients in 
the formula of $\varepsilon^{\prime\prime}$ for certain ranges of frequencies.

The goal of this study was to make measurements of the loss factor for pore water close to 
olume water for its parameters, at the frequencies $140\ldots180$~GHz and, using the 
previously obtained data, to find the analytical relation of $\varepsilon^{\prime\prime}$ 
within the frequency range from $11$ to $180$~GHz and within the temperature range from 
$0$ to $-90\,^{\circ}$C.

Among the papers published, the model proposed in \cite{MWen2004} seems to be quite suitable for low 
temperatures, as it makes use of the concept of singularity at $-45\,^{\circ}$C. 
Therefore, it was used for comparison with the new results.

\section*{Experiment}

\subsection{Analysis of the techniques of measuring the dielectric characteristics of pore 
water}

As shown in \cite{BOSch2017, BOKh2017}, the main causes of the errors made in using porous media for obtaining 
data on dielectric parameters of supercooled pore water close to volume water for its 
properties are the following: 

1. The absence of data on the fraction of the pore water close to volume water for its 
properties in the total amount of pore water.

2. Slow freezing of the pore water in a certain range of temperatures without a sharp phase 
transition, which makes determining the freezing point of the pore water difficult.

3. The emergence of cryogenic structures and of the related spatial dispersion in the medium. 
Spatial dispersion emerges due to arising inhomogeneity of the medium caused by migration of 
water to the freezing areas with the temperature gradient. 
At that, the microwave characteristics of the medium essentially change. 
	
4. The structures emerging at the freezing of the medium may display special electric 
properties. 
An example is emergence of percolation due to cross-cutting electric conductivity. 
For example, percolation was observed at microwave measurements of wetted sand at the 
frequency of $13$~GHz as its wetness changed \cite{BO2011}. 

It seems that the totality of these causes did not allow the researchers previously to use 
porous materials to measure the parameters of deeply supercooled water. 
As a result of our studies, a conclusion was made that, in order to eliminate the effects 
of the cryogenic structure and of percolation in studying the properties of supercooled pore 
water, nanoporous materials should be used with a low degree of wetness (with the moisture 
content $\sim3\ldots5$\%). 
In this case, there is practically no water in the space among the granules of the porous 
medium, which rules out migration of moisture and formation of inhomogeneities.

As noted in \cite{SchKF2001, CMSVX2016} water gets easily supercooled in small pores. 
Thus, media with a low own loss factor, like silicates, are convenient for microwave 
measurements. 
Therefore, as porous matrix, nanoporous silicate materials, silica gels, with the average 
pore size $6-9$~nm, were used in \cite{BK2012}. 
For them, we obtained good agreement between the loss factor of pore water and the 
computational data according to formulas described in \cite{MWen2004} for volume (free) water. 
It is quite important to indicate that this conclusion is corroborated by the studies 
conducted on clusters consisting of hundreds of water molecules contained in silicate pores 
by the methods of molecular dynamics \cite{LCh2012, CMSVX2016, CGAD2009}. 
It was shown that primarily the first layer of water molecules ($\sim0.25$~nm thick) 
adsorbed on the surface of the silicate pores is strongly connected with the material of the 
silicate pore walls. 
The subsequent layers of the water molecules, unlike in other materials, have parameters 
coinciding with the parameters of volume water. 
Hence, water in the pores of silicate materials may be considered as an object having the 
properties of volume water. 
Although before the moment of the phase transition at negative temperatures the liquid pore 
water is in equilibrium state, the main volume of water in the pores of the silicate 
materials may be considered as an object having the qualities of volume (metastable) water. 
For example, for pore diameters $\sim6$~nm, a considerable part of water in the pore volume 
will be 67\% for the given case, which follows from the calculation of the volumes. 
For other materials, a special study is required, as well as measuring the thickness of the 
layer of interstitial water, in order to identify the fraction of the volume water.

Supercooling of water ${\Delta}T$ in the pores of silicate materials may be determined from 
the Gibbs-Thomson equation \cite{SchKF2001, CMSVX2016}. 
${\Delta}T=T_0+T_m=c/(R-t)$, where $T_0$ is the melting point of macroscopic samples of ice 
in free space, $T_m$ is the melting temperature of ice in the pores, 
$c=62$~(degrees$\cdot$nm); $R$ is a pore radius in nm; $t\approx0.38$~nm. 
It is also known that during the processes of water cooling and heating in pores, hysteresis 
takes place, at which the freezing temperature proves to be lower than the melting 
temperature by approximately $10\,^{\circ}$C in the pores sized $5\ldots9$~nm. 
At incomplete ($\sim30$\%) filling of the pores with water, further decrease of the water 
freezing temperature in silicate materials was observed to constitute about 
$10\,^{\circ}$C \cite{SchKF2001}. 
As a result, it was found that for pores $5\ldots9$~nm in diameter, given their volume was 
less than $30$\% complete, water can be supercooled to $-50\ldots-70\,^{\circ}$C.

As noted above, a cryogenic structure emerges in freezing wetted medium, which does not 
allow the properties of pore water to be restored using standard methods. 
That was found during measurements of the reflection coefficient from freezing disperse 
medium in waveguides depending on temperature. 
Anomalies were observed also for the shape of resonance curves of the resonators completely 
filled with wetted frozen medium. 
As clusters emerge in the medium, its properties may be also affected by formation of 
ferroelectric ice~0 at the temperatures below $-23\,^{\circ}$C. 
The existence of this ice was discovered at computer simulation \cite{RRT2014, QAS2014}, and its 
influence on the microwave properties of porous media was experimentally established in \cite{BO2017}. 
During interaction with the matrix material, ice~0 forms a thin conductive layer at the 
boundary of the media due to the sharp difference in the values of static dielectric 
permittivity ($\varepsilon_s$) \cite{KDFChA2002, KMSRAP2005}. 
Hence, at occurrence of the phase transition of water into ice~0 at temperatures below 
$-50\,^{\circ}$C, for porous medium with the pore diameter $5\ldots9$~nm, we can expect 
increment of $\varepsilon^{\prime\prime}$ due to emerging conductivity in the samples.

\subsection{Experiment}

In conducting the experiments, we used finely disperse nanoporous silicates (silica gels) 
with the moisture content $3\ldots6$\%, which corresponded to $\sim10$\% filling of the 
pore volume. 
In addition, other methods of eliminating the influence of the possible inhomogeneity of 
the structure were used, which consisted in averaging the radiation in the space and by 
the frequency. 
For this purpose, flat samples placed in free space and samples placed in long waveguide 
cells were used, as well as broad-band microwave radiation in the frequency range of 
$\sim10$\% of the average frequency. 
We measured intensity of microwave radiation which passed through the medium ($I$), by the 
change of which we determined the power attenuation coefficient ($\alpha$) without 
considering the radiation scattering: $I=I_0e^{-{\alpha}Z}$, where $I_0$ is the initial 
value of the radiation intensity, $Z$ is the layer thickness. 
From the values of $\alpha$, we determined the loss factor depending on the frequency 
$\varepsilon^{\prime\prime}(f)$ using the a priori information about the values of 
$\varepsilon^{\prime\prime}$ and $\varepsilon^{\prime}$ (the real part of the relative 
dielectric permittivity) at $0\,^{\circ}$C \cite{MWen2004}. 
The measurements were made at 16 frequencies within the range from $11$ to $180$~GHz. 
The layouts of the measurement setups are shown in Fig. 1. 

\begin{figure}
\includegraphics[width=0.9\columnwidth]{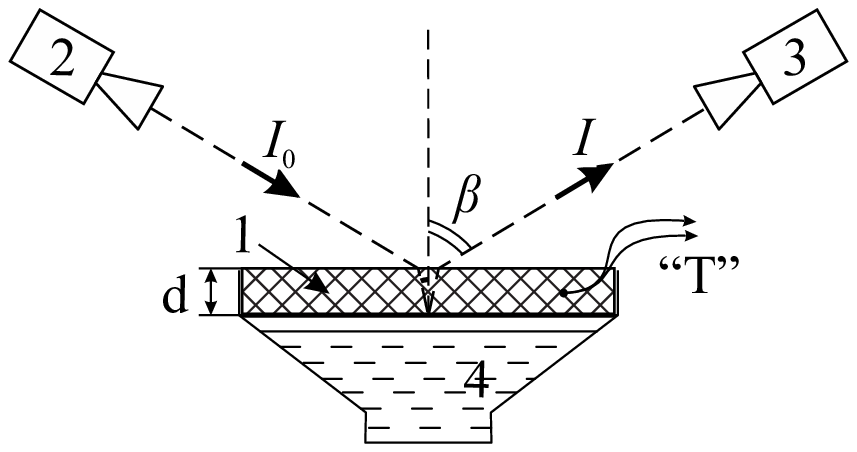}

\caption{Layouts of the measurement setups in open space. 1~--- a flat metal cuvette 
containing the sample; 2~--- a probe radiation generator; 3~--- a radiation receiver; 
4~--- a tank with liquid nitrogen; ``T''~--- thermocouples; $\beta$~--- Brewster's angle.}
\end{figure}

During the studies, the refraction model for the mixture of the porous matrix material and 
water was used. 
For this model, refractions \cite{Shar2003} were added. 
For the two-component medium:

\begin{equation}
\dot{n}=\dot{n}_1x_1+\dot{n}_2\left(1-x_1\right),
\label{eq:ncompl}
\end{equation}

where $\dot{n}$, $\dot{n}_1$, $\dot{n}_2$ are complex refractive indexes of disperse medium, 
water and matrix material, respectively; $x_1$ is the bulk concentration of water. 
Hence: 

\[
I=I_0e^{-{\alpha}_1Z_1-{\alpha}_2Z_2},
\]

where $Z_1$ and $Z_2$ are the equivalent thicknesses of water and of the matrix material, 
$\alpha_1$ and $\alpha_2$ are the respective attenuation coefficient. 
After transformation, we find $\alpha_1$:

\[
\alpha_1=\frac{1}{Z_1}\left(ln\frac{I_0}{I}-\alpha_2Z_2\right).
\]

Assuming that $\alpha_2$, $Z_1$ and $Z_2$ do not depend on temperature and do not vary in the 
given experiment, we obtain: 

\begin{equation}
\alpha_1=\frac{1}{Z_1}ln\frac{I_0}{I}-const.
\label{eq:alphaw}
\end{equation}

Disperse medium also contains air and adsorbed water (interstitial water). 
For interstitial water and for positive temperatures, the relaxation time is 
$\sim10^{-6}\ldots10^{-8}$~s, which is essentially lower than for volume water \cite{KWT2005}. 
We can expect this time to grow for interstitial water as it cools down. 
Therefore, the finite Eq. (\ref{eq:alphaw}) does not change when the refraction model is used 
for the number of components greater than two. 

For the case of measuring $\alpha$ in free space (Fig. 1(a)), in Eq. (\ref{eq:alphaw}) $Z_1$ 
should be substituted for $2d/cos\theta$, where $\theta$ is the refraction angle. 
The presented layout of measurements was used in \cite{MatWeg1987, BK1998}. 
The factor ``2'' accounts for double passing of radiation through the layer.

For measurements, a microwave radiometer with the frequency band width $\sim1\ldots3$~GHz 
was used. 
As noise generators, diode generators of microwave radiation noise were used. 
The following silicate materials were used: silica gel KSKG (made in China) and Acros 
silica gel (made in Belgium) with the average pore sizes $6\ldots9$~nm, the pore volume 
$\sim0.5$~cm$^3$/g and the pore surface area $\sim500$~m$^2$/g. 
The granules of the Acros silica gel were sized from $20$ to $50\,\mu$m, while the KSKG 
gel was crushed into granules sized $\sim100\,\mu$m. 
The values of the linear dimensions in the trough plane were chosen within 30 wavelengths, 
and it was placed in the far zone of the antennas. 
The angles of incidence and observation were chosen to be equal to Brewster's angle on 
vertical polarization. 
Due to small changes of $\varepsilon^{\prime}$ in cooling of the medium, a change in this 
angle at the fixed position of the antennas results in a measurement error 
$\alpha_1\sim1$\% due to increment in the reflection coefficient. 

\subsection{Processing of data}

In order to obtain $\varepsilon^{\prime\prime}$ from $\alpha_1$, it is necessary to 
determine the value of ``const'' in Eq. (\ref{eq:alphaw}), i.e., the contribution of the 
losses in the silicate material and in adsorbed water. 
For this purpose, we subtracted the value of 
$\frac{1}{Z_1}ln\frac{I_0}{I(-90\,^{\circ}\mathrm{C})}$ from the measured values of 
$\frac{1}{Z_1}ln\frac{I_0}{I(T)}$, as at $-90\,^{\circ}$C, liquid volume water in the 
pores turns into ice \cite{LCh2012}, and only the residual layer of adsorbed water remains. 
The obtained values of $\alpha_1^{\prime}$ turned out to be lower than the computed model 
values \cite{MWen2004} at the temperatures close to $0\,^{\circ}$C, as in determining $Z_1$ by the 
thermostatic-gravimetric method, over-estimated values were obtained of the thickness 
of the layer of water, as not all the water was free. 
To adjust $\alpha_1^{\prime}$, we used a priori information about the value of the 
attenuation factor of water at the temperature of $0\,^{\circ}$C ($\alpha_{1m}(0\,^{\circ}$C) 
obtained by calculating $\varepsilon^{\prime}$ and $\varepsilon^{\prime\prime}$ from the 
known values for liquid water \cite{MWen2004}. 
We determined a certain factor ($g$), which was multiplied by 
$\alpha_1^{\prime}(0\,^{\circ}\mathrm{C})$ to average the experimental and simulated 
values of the attenuation factors: 
$g=\alpha_{1m}(0\,^{\circ}\mathrm{C})/\alpha_1^{\prime}(0\,^{\circ}\mathrm{C})$. 
Such a method of taking adsorbed water into account was used, for example, in monograph 
\cite{Benz1974}.

As a result, the values ($\alpha_1^{\prime}{\cdot}g$) proved to agree well with the 
simulated values of $\alpha_1$ according to model \cite{MWen2004} to the temperature 
$-20\ldots-30\,^{\circ}$C. 
From the known ratios for the attenuation coefficient and $\varepsilon^{\prime}$, 
$\varepsilon^{\prime\prime}$ \cite{Shar2003} for a medium without dissipation, we found the loss 
factor of supercooled pore water $\varepsilon^{\prime\prime}$. 

\[
\alpha=\frac{4\pi\kappa}{\lambda_0}; \dot{n}=\sqrt{\dot{\varepsilon}}=n+i\kappa; 
\kappa=\sqrt{0.5}\left[\sqrt{\varepsilon^{\prime2}+\varepsilon^{\prime\prime2}}-\varepsilon^{\prime}\right]^{1/2},
\]

where $\lambda_0$ is the wavelength in free space; $\kappa$ is the imaginary part of 
the complex refractive index.

In finding $\varepsilon^{\prime\prime}$ the values of $\varepsilon^{\prime}$ were used 
from \cite{MWen2004} in the temperature range of $0\ldots-40\,^{\circ}$C. 
In the temperature range of $-40\ldots-90\,^{\circ}$C, the value from $2.3$ to $3.15$ 
(depending on frequency) was used, in accordance with the calculation from \cite{MWen2004} for the 
temperature of $-44.5\,^{\circ}$C.

The assumed approximation was verified, as the values of $\varepsilon^{\prime}$ at the 
temperatures below $-20\,^{\circ}$C did not have experimental confirmation. 
For this purpose, the errors of determining $\varepsilon^{\prime\prime}$ at changing 
$\varepsilon^{\prime}$ were assessed. 
The used values of $\varepsilon^{\prime}$ were changed two times from $3$ to $6$ at 
the frequencies from $15$~GHz to $120$~GHz and in the temperature range from 
$-30$ to $-50\,^{\circ}$C. 
The minimum calculated value $\Delta\varepsilon^{\prime\prime}/\varepsilon^{\prime\prime}$ 
was $18$\% at $15$~GHz at $-30\,^{\circ}$C. 
At the same frequency and at the temperature of $-50\,^{\circ}$C the operational margin 
reached $27$\%. 
As the frequency rose to $120$~GHz, the operational margin, with the used error in the 
knowledge of $\varepsilon^{\prime}$, gradually rose to constitute 
$34$\% at $-30\,^{\circ}$C and $36$\% at $-50\,^{\circ}$C. 
However, possible deviations of $\varepsilon^{\prime}$ from the used values are not so 
high, therefore, the expected errors of determining $\varepsilon^{\prime\prime}$ are to 
be less by a factor of $2-3$, compared to the estimations provided. 

The operational margin of the relative measurements of $\alpha$ of the medium by 
Eq. (\ref{eq:alphaw}) was $15$\%, and that for $\varepsilon^{\prime\prime}\sim20$\% at 
lower frequencies at $-30\ldots-40\,^{\circ}$C. 
As temperature decreased to $-50\,^{\circ}$C, the operational margin of measuring 
$\varepsilon^{\prime\prime}$ rose to $25$\%. 
In the upper part of the studied range of temperatures, this error was $25$\% for 
$-30\,^{\circ}$C and $30$\% for $-50\,^{\circ}$C. 
The operational margin of measuring $\varepsilon^{\prime\prime}$ at the temperatures 
below $-50\,^{\circ}$C rose due to transition of a part of the water into 
a solid state. 
To reduce the errors of determining $\varepsilon^{\prime\prime}$, it is necessary 
to measure $\varepsilon^{\prime}$, which is an specific task. 
However, this range of temperatures requires a special study.

\subsection{The measurement results}

The results of determining $\varepsilon^{\prime\prime}$ depending on frequency for 
several frequencies are provided in Fig. 2(a, b, c).

\begin{figure}
\includegraphics[width=0.9\columnwidth]{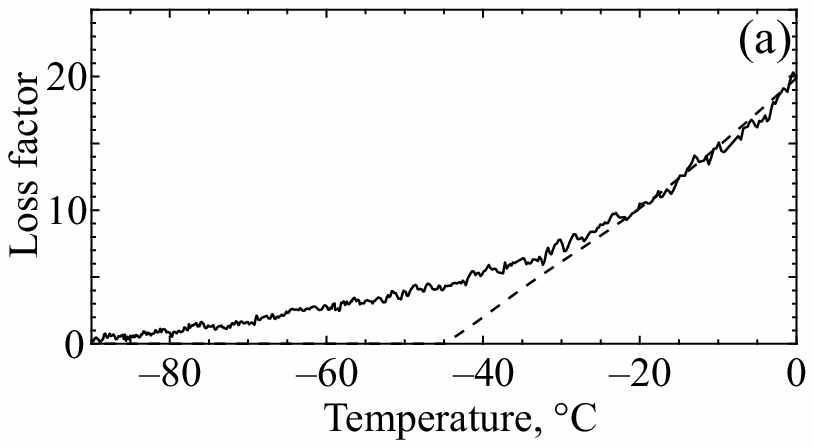}
\includegraphics[width=0.9\columnwidth]{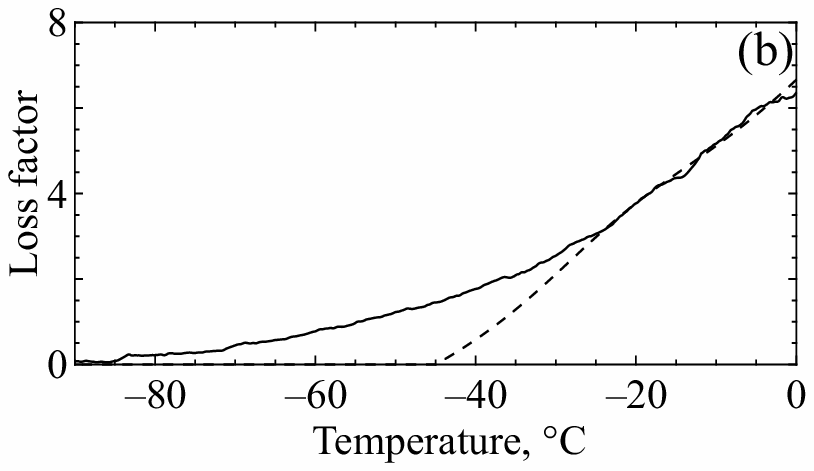}
\includegraphics[width=0.9\columnwidth]{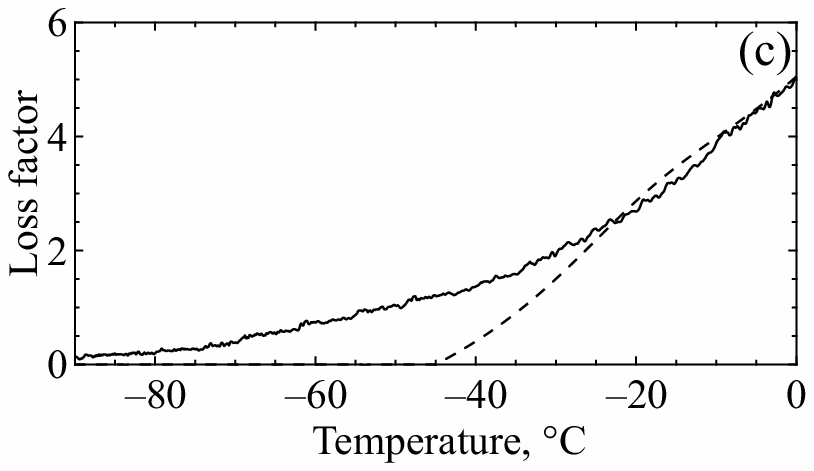}

\caption{The results of determining $\varepsilon^{\prime\prime}$ of supercooled pore 
water close to volume water for its properties at the frequencies: a) $34$~GHz; 
b) $122$~GHz; c) $175$~GHz. 
The solid line presents the experimental data; the dashed line represents the modeled 
values, in accordance with \cite{MWen2004}.}
\end{figure}

Altogether, we made measurements for $50$ different samples within the range of 
frequencies from $11$~GHz to $180$~GHz and the range of temperatures 
$0\ldots-90\,^{\circ}$C. 
As it follows from the measurements, in the range of temperatures below $-30\,^{\circ}$C, 
significant difference between the obtained data and the simulation results was observed. 
We found the difference $\Delta\varepsilon^{\prime\prime}$ between the experimental 
values $\varepsilon^{\prime\prime}$ and the model values $\varepsilon_m^{\prime\prime}$ 
according to model \cite{MWen2004}, which has a look close to a bell-shaped curve with a 
characteristic extremum near $-45\,^{\circ}$C and certain asymmetry. 
The diagram of $\Delta\varepsilon^{\prime\prime}$ of dependence on temperature for 
frequency of $34$~GHz is shown in Fig. 3. 

\begin{figure}
\includegraphics[width=0.9\columnwidth]{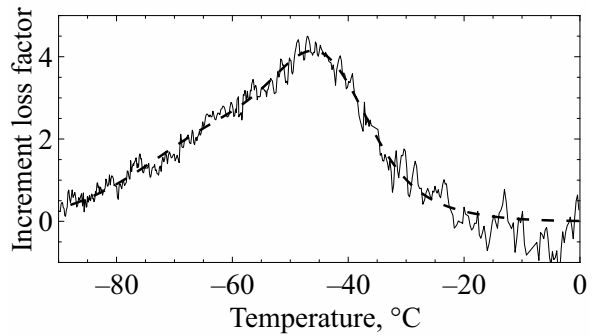}

\caption{The difference between the values of measured and calculated 
$\varepsilon^{\prime\prime}$ for pore water close to volume water for its properties 
and volume water according to model \cite{MWen2004} depending on temperature at the frequency 
of $34$~GHz. 
The smooth dashed curve is approximation by a sum of two Gaussian functions.}
\end{figure} 

Additionally, $\varepsilon_m^{\prime\prime}$ was considered equal to zero in the 
model \cite{MWen2004} at $T\leq-45\,^{\circ}$C.

The curve $\varepsilon_m^{\prime\prime}$ is well approximated by a sum of two Gaussian 
functions in the temperature range $-20\ldots-90\,^{\circ}$C.

\begin{equation}
\Delta\varepsilon^{\prime\prime}=a_1e^{-\left(\frac{T-T_1}{c_1}\right)^2}+a_2e^{-\left(\frac{T-T_2}{c_2}\right)^2}, 
T\leq0\,^{\circ}\mathrm{C},
\label{eq:depsilon}
\end{equation}

where $T$ is in Centigrade degrees. 
$T_1=-45\,^{\circ}$C, for $T_2$ the value of $-60\,^{\circ}$C was used. 
In the general case, $T_1$ within one degree matched $-45\,^{\circ}$C, $T_2$ varied 
within $-60\ldots-70\,^{\circ}$C.

To ensure analytical description of $\Delta\varepsilon^{\prime\prime}$ depending not 
only on temperatures but also on frequency we found, by way of optimization, 
dependences $a_1$, $c_1$, $a_2$, $c_2$ in Eq. (\ref{eq:depsilon}) for the frequency 
range of $11\ldots180$~GHz. 
The functions which are closest to the experimental data are the following:

\[
a_1=10.91e^{-0.1267f}+2.672e^{-4.777\times10^{-3}f}; 
\]

\[
c_1=1.066\times10^{-6}f^3-6.52\times10^{-4}f^2+0.1293f+7.779; 
\]

\[
a_2=4.16e^{-0.0101f}; 
\]

\begin{equation}
c_2=2.873\times10^{-6}f^3-6.945\times10^{-5}f^2-7.64\times10^{-3}f+15.4,
\label{eq:coeff}
\end{equation}

where $f$ is in GHz. 
$T_1=-45\,^{\circ}$C, $T_2=-60\,^{\circ}$C.

Thus, the loss factor of supercooled water within the range of temperatures 
$0\ldots-90\,^{\circ}$C was shown as:

\begin{equation}
\varepsilon^{\prime\prime}=\varepsilon_m^{\prime\prime}+\Delta\varepsilon^{\prime\prime}, 
(\varepsilon_m^{\prime\prime}=0\ \mathrm{at}\ T\leq-45\,^{\circ}\mathrm{C}).
\label{eq:epsilon}
\end{equation}

The results of calculations for $\varepsilon^{\prime\prime}$ depending on frequency 
($f$) and of several values of temperature are shown in Fig. 4.

\begin{figure}
\includegraphics[width=0.9\columnwidth]{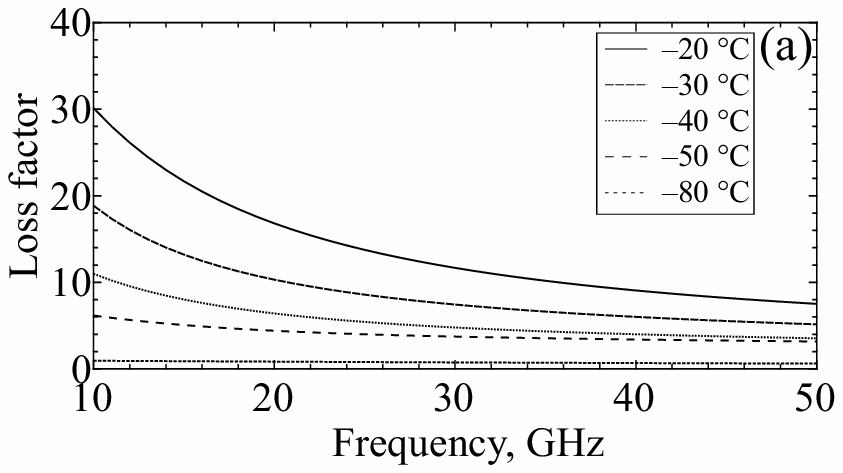}
\includegraphics[width=0.9\columnwidth]{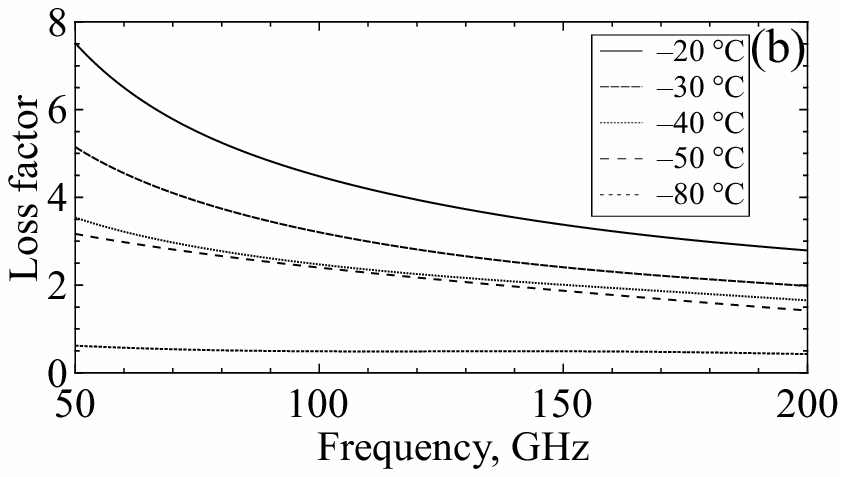}

\caption{The loss factor of supercooled water in the frequency ranges: a)~--- $10\ldots50$~GHz; 
b)~--- $50\ldots200$~GHz, determined by Eq. (\ref{eq:depsilon} -- \ref{eq:epsilon}). }
\end{figure}
 
Shown in Fig. 5 are the curves of dependences $\Delta\varepsilon^{\prime\prime}$ on 
temperature for three values of frequencies.

\begin{figure}
\includegraphics[width=0.9\columnwidth]{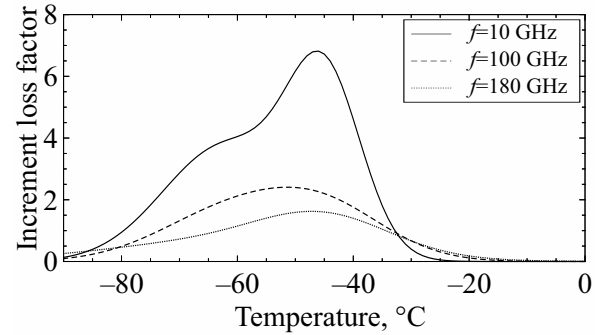}

\caption{Diagrams of the values of the values of $\Delta\varepsilon^{\prime\prime}$ in the 
formula of the loss factor of supercooled water within the range of temperatures 
$0\ldots-90\,^{\circ}$C for three values of frequencies determined by 
Eq. (\ref{eq:depsilon}, \ref{eq:coeff}).}
\end{figure}

\section*{Discussion of results}

It is to be noted that volume water does not exist below $-45\,^{\circ}$C \cite{FHKMSSSS2009}. 
In the experiment described in \cite{KSPPM2017}, minimum temperature of $-44\,^{\circ}$C was achieved 
for nano drops within microseconds. 
In clouds the temperature $-37.5\,^{\circ}$C was observed \cite{RosWood2000}, corresponding to the 
temperature range of ``no man's land'' ($-37\ldots-120\,^{\circ}$C). 
Therefore, the obtained results of microwave measurements for the temperature of 
$-45\,^{\circ}$C and below correspond to pore water close to volume water for its properties, 
but in which formation of nucleating centers is suppressed due to interaction between the 
water and the pore walls. 
To a certain extent, this effect should take place also in micro emulsions, for which 
measurements were made in \cite{BCS1982}, and also in the micro droplets liquid in the atmosphere. 

On the other hand, these data for the range of temperatures $-45\ldots-90\,^{\circ}$C are 
also of certain interest due to the possibility of deep supercooling of water in hard 
porous aerosol particles in the atmosphere with nanopores. 
Unfrozen water also exists at these temperatures in soils, vegetation cover, and other 
porous media.

Analyzing the experimental data, we found out that in all the experiments the parameter 
$T_1$ for two brands of silica gel is close to the value of $-45\,^{\circ}$C within the 
temperature range of about $1\,^{\circ}$C. 
This is not related to the special shape of functions $\varepsilon_m^{\prime\prime}$ and its 
kink, as we observed a weak extremum near $-45\,^{\circ}$C in the original data of a number 
of experiments for $\alpha_1$. 
In most cases, parameter $T_2$ was within $-60\ldots-70\,^{\circ}$C. 
In formula (\ref{eq:depsilon}) it was taken to be equal to $-60\,^{\circ}$C. 
In the experiments using the silicate material SBA-15, the extremum of losses at deep 
supercooling of water was revealed at $-80\,^{\circ}$C. 
Granted that two addends in (\ref{eq:depsilon}) with extremum values at certain temperatures 
correspond to two mechanisms of electromagnetic losses, the first addend refers to properties 
manifested by purely volume water, while the second addend is related to the influence of the 
pore space (the pore material, the character of wetting, the geometry of pores, and the 
thermal history of the sample). 

The specifics of electromagnetic losses of supercooled water in pores determined by the first 
addend in (\ref{eq:depsilon}) at $-45\,^{\circ}$C may be related to the existence of the 
second critical point of water at the temperature of $-53\,^{\circ}$C and the pressure 
$30\ldots100$~MPa \cite{A2012}. 
In accordance with \cite{W1963}, for the case of the liquid-liquid critical point, in the phase diagram 
a special line named the Widom line come from it. 
In this line, fluctuations of entropy and water density rise sharply. 
At the pressure of $0.1$~MPa, the temperature in the Widom line for water is equal to 
$-45\,^{\circ}$C \cite{A2012, FSt2007}. 
In the experiments conducted on measuring the thermal capacity at constant pressure, isothermal 
compressibility and several other values to the temperature of $-39\,^{\circ}$C \cite{ASO1982}, sharp 
increase of those values was discovered at approximation to $-45\,^{\circ}$C, reminding of the 
singularity. 
For the case of pore water, for which deeper supercooling is possible, extremum with the end 
value of at $-45\,^{\circ}$C were discovered for nanopores, but not singularity \cite{NKON2010}. 

The increase of $\varepsilon^{\prime\prime}$ for supercooled pore water at this temperature 
and in a certain range of temperatures may be attributed to the increase in the entropy 
fluctuations, i.e., to fluctuations of the structure of the dipole liquid.

The second addend in (\ref{eq:depsilon}) may be related to the result of interaction between 
the molecules of water with the material of the pore surface. 
In \cite{KDFChA2002, KMSRAP2005} emergence of a conductive layer $h\sim0.1\ldots1$~nm thick on the boundary of 
dielectrics with a large difference of static dielectric permittivity values ($\varepsilon_s$). 
Its conductivity may rise $N\sim10^5$~times, compared to boundary media for the contrast 
$\Delta\varepsilon_s\sim100\ldots1000$. 
The degree of the influence of this layer on $\varepsilon^{\prime\prime}$ is determined by the 
area of the surface of the media boundary in a unit of volume. 

The recent theoretical studies of water freezing stated the emergence of ferroelectric ice 
named ice~0, which may be formed only from supercooled water, followed by formation of ice~Ih 
or Ic \cite{RRT2014, QAS2014}. 
This process may take place, according to the cited studies, at temperatures below 
$-23\,^{\circ}$C. 
Previously the rise of $\varepsilon_s$ at the temperature below $-37\,^{\circ}$C was discovered 
for water in silicate material MCM-41 with cylindrical pores \cite{FMBO2011} at low frequencies of the 
order of $10$~Hz.

The increment $\Delta\varepsilon^{\prime\prime}$ for this case may be found, knowing the 
specific conductivity ($\sigma_0$) of wetted silicate, determined by the contact layer on the 
boundary. 
$\Delta\varepsilon^{\prime\prime}\approx\left(N_0\right)/\left(\varepsilon_0\omega\right)$, 
here $\varepsilon_0$ is the electric constant, $\omega$ is the angular frequency of the 
electromagnetic field. $\sigma_0$ is calculated from the fraction of the volume occupied by the 
conductive layer with thickness $h$, as well as by area $S$ of the pore surface in a unit of 
volume and $\sigma_w$~--- by the conductivity of volume water. 
If $\sigma_w=10^{-3}\ldots10^{-4}$~Ohm$^{-1}$m$^{-1}$; $N=10^5$; $S\sim100$~m$^2$, the 
estimation provides $\Delta\varepsilon^{\prime\prime}$ at the frequency of $10$~GHz $\sim10$, 
and at the frequency $100$~GHz $\sim1$ \cite{BOKh2017}. 
Thus, the computation demonstrates that the increase $\varepsilon^{\prime\prime}$ at the 
temperatures of $-37\ldots-90\,^{\circ}$C may be explained by the emergence of the ferroelectric 
state. 
This mechanism was experimentally tested in \cite{BO2017} by three different methods: measuring of 
reflection coefficient from wetted nanoporous silicate, measuring low-frequency dielectric 
permittivity and hysteresis of electric noise; emergence of anomalies was confirmed for the 
temperatures below $-23\,^{\circ}$C.

In practical application of Eq. (\ref{eq:depsilon}) for calculating the loss factor at 
temperatures below $-60\,^{\circ}$C, it is required to account for the fraction of water which 
turned into ice and to know parameter $T_2$, at which the influence of ferroelectric ice and 
the increase of surface conductivity at the contact of media are observed. 
This parameter may vary within $-23\ldots-90\,^{\circ}$C, both depending on $\varepsilon_s$ of 
the media, the chemical composition of pores and on conductivity $\sigma_w$, which depend on the 
temperature and admixtures in water. 
It may be likely that for water aerosol this effect is insignificant; therefore, in order to 
determine attenuation of microwave radiation in the atmosphere when (aerosol) particles are 
cooled down to $-42\ldots-44\,^{\circ}$C, it is sufficient to use only the first addend in 
Eq. (\ref{eq:depsilon}).

\section*{Conclusions}

1. Under certain conditions, pore water in nanoporous silicate materials is close to volume 
water for its properties, which allows its use as a model of medium in experimental studies 
of supercooled volume water. 
An additional supposition consists in the fact that the monomolecular adsorbed layer of water 
on the pore surface does not affect the microwave parameters of pore water in such materials 
at the frequencies higher than $11$~GHz. 
It is only necessary to consider its volume fraction in the total mass of water.

2. To find the microwave parameters of volume water from the measurements of wetted nanoporous 
silicates, a special technique is required. 
It consists in the following. 
In freezing of disperse medium, the texture emerges, having macroscopically inhomogeneous 
electric characteristics. 
These inhomogeneities essentially change the microwave properties of materials due to the 
effects of spatial dispersion, which does not allow the parameters of individual components 
of the medium to be determined. 
To eliminate the influence of inhomogeneity in determining the characteristics of pore water, 
a method of measurements was used in averaging the results in space and in a certain band 
of frequencies. 
In addition, to rule out the influence of structure inhomogeneity emerging from migration of 
water, measurements were conducted on specimens of disperse porous material with low moisture 
content. 

3. It was established from the results of determining the attenuation coefficient and the loss 
factor of pore water close to volume water for its characteristics that there is additional 
increase of $\alpha_1$ and $\varepsilon^{\prime\prime}$ at the temperatures below $-30\,^{\circ}$C 
within the frequency range of $11\ldots180$~GHz. 
In \cite{MWen2004}, these values were close to zero at the temperatures below $-45\,^{\circ}$C, which the 
authors related to impossibility of the existence of liquid water at these temperatures. 
However, liquid water may exist in solid porous bodies at the temperatures below $-45\,^{\circ}$C.
 
4. Analysis of the experimental data resulted in the proposal of two mechanisms raising the pore 
water loss factor at the temperatures below $-30\,^{\circ}$C. 
The first mechanism is determined by microwave absorption in volume water, while the second 
mechanism is attributed to the surface effect of rising conductivity at the boundary of medium 
partition. 
Extremum $\Delta\varepsilon^{\prime\prime}$ at $-45\,^{\circ}$C may be explained by the influence 
of the second critical point of water on its properties existing at the temperature of 
$-53\,^{\circ}$C and the pressure of $30\ldots100$~MPa. 
The Widom lines go from this point in the phase diagram, corresponding to the temperature of 
$-45\,^{\circ}$C at atmospheric pressure. 
In the Widom line, the theoretically predicted intense entropy and water density fluctuations 
were discovered. 
The microwave measurements of absorption of the radiation of wetted porous medium support the 
existence of the second critical point of water. 

In the range of temperatures near $-60\ldots-70\,^{\circ}$C for nanoporous silicates, we found 
increase $\varepsilon^{\prime\prime}$, which we related to the specifics of water freezing at 
its supercooling below $-23\,^{\circ}$C. 
In this case, formation of ferroelectric ice~0 is possible. 
A thin layer of increased conductivity emerges at the boundary between such ice and the pore 
surface due to the high values of the difference in the static dielectric permittivity. 
The influence of such layers depending on the type of material and the effects of emergence of 
ferroelectricity requires further investigations. 

5. The microwave properties of supercooled water are of practical interest for determining the 
parameters of certain objects of the cryosphere at remote sensing. 
In addition to the known problem of studying cloud formations in the atmosphere, we can mention 
the problem of the transfer of radiation in solid aerosol particles. 
The same is true of soils, of snow and vegetation covers. 
Radiospectroscopy of wetted media may be used to study the structure of porous dielectrics based 
on the values of the loss factor and the phase transitions in them.

\bibliographystyle{apsrev4-1}
\bibliography{LGC_Bib}

\end{document}